\documentclass[brazilian,twocolumn,prb,superscriptaddress,amsmath,amssymb,showpacs,floatfix,preprintnumbers]{revtex4-2}
\usepackage[latin9,utf8]{inputenc}
\usepackage{float}
\usepackage{textcomp}
\usepackage{amsmath}
\usepackage{graphicx}
\makeatletter
\usepackage{url}
\usepackage{grffile}
\usepackage{bbold}
\usepackage{comment}
\usepackage{dirtytalk}
\usepackage{chngcntr}

\newcommand{\lyxmathsym}[1]{\ifmmode\begingroup\def\b@ld{bold}
  \text{\ifx\math@version\b@ld\bfseries\fi#1}\endgroup\else#1\fi}

\@ifundefined{textcolor}{}
{%
 \definecolor{BLACK}{gray}{0}
 \definecolor{WHITE}{gray}{1}
 \definecolor{RED}{rgb}{1,0,0}
 \definecolor{GREEN}{rgb}{0,1,0}
 \definecolor{BLUE}{rgb}{0,0,1}
 \definecolor{CYAN}{cmyk}{1,0,0,0}
 \definecolor{MAGENTA}{cmyk}{0,1,0,0}
 \definecolor{YELLOW}{cmyk}{0,0,1,0}
}

\usepackage{dcolumn}
\usepackage{color}
\DeclareGraphicsExtensions{.png .jpg .pdf}

\usepackage{hyperref}
\hypersetup{
     colorlinks = true,
     linkcolor = blue,
     anchorcolor = blue,
     citecolor = blue,
     filecolor = blue,
     urlcolor = blue
     }
\usepackage{braket}
\usepackage{physics}
\usepackage{array}
\usepackage{booktabs}
\usepackage{soul}

\makeatother

\usepackage[printwatermark]{xwatermark}
\usepackage{xcolor}
\usepackage{graphicx}
\usepackage{tikz}

\newsavebox\mybox
\savebox\mybox{\tikz[color=gray,opacity=0.4]\node{arXiv Version};}
\newwatermark*[
  allpages,
  angle=45,
  scale=12,
  xpos=-50,
  ypos=50
]{\usebox\mybox}

\begin{document}
\title{Tunable coupling of terahertz Dirac plasmons and phonons in transition metal dichalcogenide-based van der Waals heterostructures}
\author{I. R. Lavor}
\email{icaro@fisica.ufc.br}
\affiliation{Instituto Federal de Educação, Ciência e Tecnologia do Maranhão, KM-04,
Enseada, 65200-000, Pinheiro, Maranhão, Brazil}
\affiliation{Departamento de F\'{i}sica, Universidade Federal do Ceará, Caixa Postal
6030, Campus do Pici, 60455-900 Fortaleza, Ceará, Brazil}
\affiliation{Department of Physics, University of Antwerp, Groenenborgerlaan 171,
B-2020 Antwerp, Belgium}
\author{Andrey Chaves}
\affiliation{Departamento de F\'{i}sica, Universidade Federal do Ceará, Caixa Postal
6030, Campus do Pici, 60455-900 Fortaleza, Ceará, Brazil}
\affiliation{Department of Physics, University of Antwerp, Groenenborgerlaan 171,
B-2020 Antwerp, Belgium}
\author{F. M. Peeters}
\affiliation{Department of Physics, University of Antwerp, Groenenborgerlaan 171,
B-2020 Antwerp, Belgium}
\author{B. Van Duppen}
\email{ben.vanduppen@uantwerpen.be}
\affiliation{Department of Physics, University of Antwerp, Groenenborgerlaan 171,
B-2020 Antwerp, Belgium}
\date{\today }
\begin{abstract}
Dirac plasmons in graphene hybridize with phonons of transition metal dichalcogenides (TMDs) when the materials are combined in so-called van der Waals heterostructures (vdWh), thus forming surface plasmon-phonon polaritons (SPPPs). The extend to which these modes are coupled depends on the TMD composition and structure, but also on the plasmons' properties. By performing realistic simulations that account for the contribution of each layer of the vdWh separately, we calculate how the strength of plasmon-phonon coupling depends on the number and composition of TMD layers, on the graphene Fermi energy and the specific phonon mode. From this, we present a semiclassical theory that is capable of capturing all relevant characteristics of the SPPPs. We find that it is possible to realize both strong and ultra-strong coupling regimes by tuning graphene's Fermi energy and changing TMD layer number. 
\end{abstract}
%%\pacs{XXXXX-XXX-XX}
\maketitle

\section{Introduction}\label{Sec. Intro}

In the past few years, after the advent of graphene~\citep{Novoselov2004}, a two-dimensional (2D) monolayer of carbon atoms arranged in a honeycomb lattice, the interest of the scientific community in isolating and studying new 2D materials has been significantly increasing due to the unique features of these materials~\cite{Butler2013,Geim2013,Fiori2014}.  %One of the reasons, is that,%
For example, 2D transition metal dichalcogenides (TMDs)~\citep{Manzeli2017,Novoselov2005}, such as MoS$_2$, MoSe$_2$, WS$_2$ and WSe$_2$, have attracted considerable attention due to their remarkable opto-electronic properties~\citep{Geim2013,Wang2012,Low2014,Low2016,Ranieri2014,Jariwala2014,Zhang2015,Ju2011,Chen2012,Fiori2014,Mak2016} that arises, for example, due to their electronic band gaps~\citep{Yu2015,Mak2010}, the specific type of the electronic structure, and the intrinsic mobility of the electrons~\citep{Neto2009}. These 2D materials can be combined in so-called van der Waals heterostructures (vdWh)~\citep{Geim2013,Novoselov2016,Liu2016} by stacking different layers on top of each other~\citep{Geim2013,Wang2012,Jariwala2014,Zhang2015,Mak2016,gong2014vertical,Manzeli2017}, or even next to each other forming so-called lateral heterostructures~\cite{ozcelik2016band,sahoo2018one,duan2014lateral,gong2014vertical,gong2015two,huang2014lateral,Manzeli2017}, resulting in the creation of many different multi-layered artificial materials, each with specific behaviour~\citep{Liu2016,Jariwala2016}. Recently, significant advances have been made to obtain and manufacture such heterostructures~\cite{Jariwala2016,ozcelik2016band,sahoo2018one,duan2014lateral,gong2014vertical,gong2015two,huang2014lateral,Geim2013,Wang2012,Novoselov2005,Liu2016,Zhang2015,Manzeli2017}.

Graphene plasmons, collective excitations of the 2D electron liquid in graphene~\citep{GabrieleGiuliani2008,Maier2007}, also known as Dirac plasmons~\citep{Grigorenko2012}, are heavily studied due to their low loss~\cite{Yan2013,Jablan2009}, a frequency that is tunable by the Fermi energy~\cite{Chen2012,Fei2012,Ju2011,Fei2011} and their possible applications in photonics~\cite{Cai2014,Grigorenko2012,Liu2013}. Besides, graphene can support plasmons at mid infrared (IR) \citep{Zhong2015,Schuller2010,Low2014} to terahertz (THz) frequencies~\citep{Polini2016,Low2014,Ju2011,Alonso-Gonzalez2016} and show strong electromagnetic field confinement~\citep{Goncalves2015,Grigorenko2012}. On the other hand, in TMDs (such as MoS$_2$ or WS$_2$, for example), active modes reside in the mid-IR range~\cite{Zhang2015a} and, due to their large electronic band gap~\citep{Mak2010,Splendiani2010}, these materials behave as dielectrics at low frequencies, thus not supporting plasmons if not extrinsically doped~\cite{Li2015}.

\begin{figure}[!t]
\centering{}\includegraphics[width=1\columnwidth]{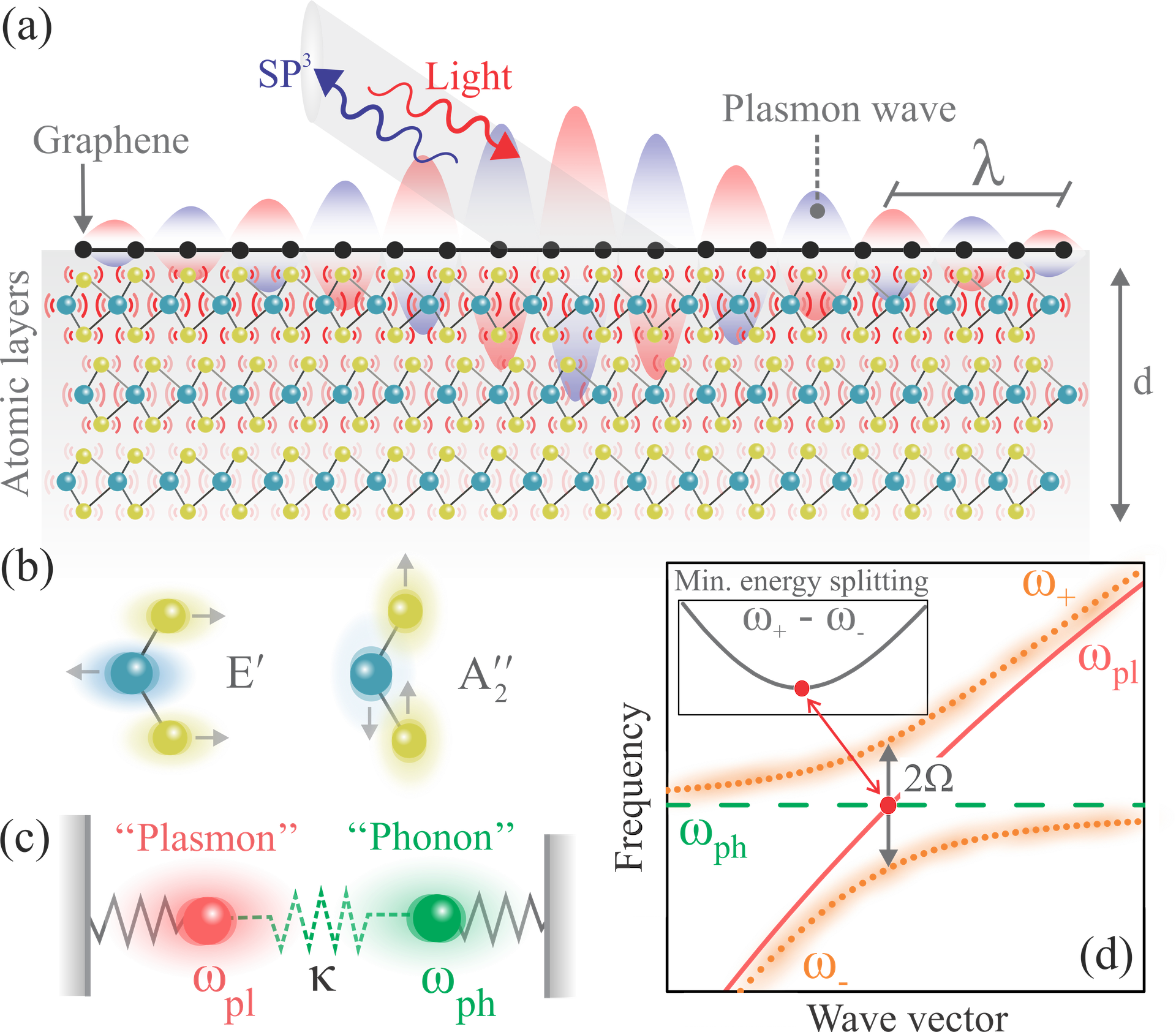}
\caption{(Color online) (a) Schematic illustration of the Dirac plasmon wave and the phonon-polariton vibration in van der Waals heterostructures (vdWh) composed by a monolayer graphene (G) on $\text{3-MX}_{2}$ (M=W,Mo and X=S,Se). The graphene surface plasmon-phonon polariton wavelength is $\lambda$. Note that the monolayer graphene covers the entire sample. The hybridization of the phonon-polariton vibration in a vdWh with the Dirac plasmon  originates from the hybridized surface plasmons ($\text{SP}_{3}$). (b) Representation of the in-plane (E$_{\prime\prime}$) and out-of-plane (A$^{\prime\prime}_2$) phonon vibration. (c) Plasmon and phonon coupling pictorially depicted as two coupled classical mechanical oscillators. The strength of the coupling is determined by $\kappa$ and gives rise to a splitting in the two eigenfrequencies $\omega_{ph}$ and $\omega_{pl}$. (d) Qualitative representation of the eigenfrequencies $\omega_{ph}$ (horizontal green dashed line) and $\omega_{pl}$ (solid red line) of the uncoupled ($\kappa=0$) plasmon-phonon system. The modes of the coupled system is represented by the upper ($\omega_+$) and lower ($\omega_-$) eigenfrequencies (orange dotted lines), its difference is called minimal energy splitting (see inset). $\Omega$ quantifies the strength of the plasmon-phonon coupling.}
\label{Fig_vdWhs_plamon_phonon_coupling}
\end{figure}

As illustrated in Fig.~\ref{Fig_vdWhs_plamon_phonon_coupling}(a), when a monolayer graphene (MLG) is combined with layers of TMDs, forming graphene-based vdWhs, a hybrid excitation arises that is known as surface plasmon-phonon polaritons (SPPPs). These quasiparticles are formed when phonons in the TMDs are coupled to the electron oscillations in graphene~\cite{Goncalves2015}. One can excite and measure them using scatter-type scanning near-field optical microscopy (s-SNOM)~\citep{Lundeberg2017,Lundeberg2016,Alonso-Gonzalez2016}. This allows one to measure the SPPPs wavelength, with a resolution of up to $20$ nm~\citep{Woessner2014,Fei2012,Fei2015,Lundeberg2017,Lundeberg2016,Dai2015,Chen2012,Dai2014}, using interference fringes formed by the scattering of SPPPs modes at the edge of the heterostructure or at lateral defects in the system. Although monolayer TMDs have four phonon modes in the IR spectrum, only two of them, the in-plane $\text{E}^{\prime\prime}$ and out-of-plane $\text{A}^{\prime\prime}_2$, illustrated in Fig.~\ref{Fig_vdWhs_plamon_phonon_coupling}(b), are IR-active and can couple to Dirac plasmons~\cite{Hertzog2019,Xia2017}.

In analogy to two coupled harmonic oscillators~\cite{Novotny2010}, Fig.~\ref{Fig_vdWhs_plamon_phonon_coupling}(c), when graphene plasmons and TMDs phonons are coupled, the eigenfrequencies of the system are modified, presenting a characteristic anti-crossing~\cite{Hertzog2019,torma2014strong}, as shown in Fig.~\ref{Fig_vdWhs_plamon_phonon_coupling}(d). By investigating the specific way in which the anti-crossing is manifested, one can infer the way in which hybridization occurs, quantified by the coupling strength, $\Omega$, between Dirac plasmons and environmental phonons. 

Whether hybridization is significant or not depends on the strength of the plasmon-phonon coupling when compared to other relevant energy scales, for example, the phonon energy and linewidth~\cite{torma2014strong}. The latter is schematically presented in Fig.~\ref{Fig_vdWhs_plamon_phonon_coupling}(d) as an orange shade along the hybrid modes. In this context, the splitting becomes only significant when the coupling $\Omega$ exceeds the linewidths of the two coupled systems, which also enables the experimental observation of these two modes. Thus, if $\Omega$ is very small compared to other important energy scales, for example, the phonon energy, the coupling is negligible and is not strong enough to change the original (uncoupled) frequencies. This defines different coupling regimes: the first one, where $\Omega$ is small, is classified as "weak coupling" (WC)~\cite{torma2014strong,Bitton2019}. On the other hand, if $\Omega$ is large when compared to the phonon energy, the coupling modifies the original energy spectrum, creating hybrid plasmon-phonon modes. In this case, the coupling regime is classified as "strong" (SC) or "ultrastrong" coupling (USC)~\cite{torma2014strong,Bitton2019,comment1}. The latter enables more efficient plasmon-phonon interactions, resulting in electro-optical devices with high efficiency when compared to those based on SC~\cite{kockum2019ultrastrong}. For the purposes of this article, we define the WC, SC and USC regimes in a pragmatic way: after obtaining $\Omega$, we normalize the coupling strength in relation to the phonon frequency that originates the hybridization as $\eta = \Omega / \omega_{ph}$; then, we classify the system as WC, SC and USC when $\eta<0.01$, $0.01 \leq \eta < 0.1$ and $ \eta \geq 0.1$, respectively~\cite{kockum2019ultrastrong}.

In this paper an investigation on the coupling between Dirac plasmon and IR-active TMDs phonons is presented. Through realistic simulations at the level of density functional theory (DFT), many-body perturbation theory and the random phase approximation (RPA)~\cite{Haastrup2018,gjerding2021recent}, in combination with the quantum electrostatic heterostructure model (QEH)~\citep{Andersen2015}, we are able to investigate the way in which the plasmon-phonon coupling depends on the number of heterostructure layers, define the coupling regime and, more significantly, identify how the Fermi energy contributes to maximize the coupling strength. Furthermore, the use of QEH also allows us to analyse how the properties of the environment are affected even when a single monolayer is added to the vdWhs. We show that a semiclassical theory within the RPA is capable of capturing all relevant characteristics of the SPPPs coupling taking into account the TMDs thickness up to several layers. Therefore, we provide a realistic evaluation of the way in which the phonon modes of the TMDs layers couple to the electromagnetic field of the plasmon modes and describe the dependence of the coupling strength up to the bulk limit. Finally, we show how controlling the graphene Fermi energy can maximize the coupling, towards SC and USC regimes in TMDs-based vdWhs. Although the study presented here considers only hexagonal MoS$_2$ and WS$_2$, it can easily be extended to all other TMDs.

The paper is structured as follows: In Sec.~\ref{sec:THEORETICAL-FRAMEWORK} we introduce the theoretical treatment of Dirac plasmons in vdWhs, by presenting an effective dynamical non-local background dielectric function that takes into account the TMDs thickness, the semi-classical RPA-based theory and the way in which the QEH calculates the role of each layer separately. In Sec.~\ref{sec:RESULTS-AND-DISCUSSIONS}, we present the results of the plasmon-phonon dispersion in the $(q ,\omega)$-plane emphasizing which phonon modes are significantly coupled to the graphene plasmons and compare the QEH results to those from the semi-classical model. Then, we show the dependence of the SPPPs coupling on the number of TMDs layers for the IR-active in-plane ($\text{E}^{\prime}$) and out-of-plane ($\text{A}^{\prime\prime}_2$) phonon modes, highlighting which of them are in WC, SC, or USC regimes, through the normalization of the coupling $\eta$. By the end of Sec.~\ref{sec:RESULTS-AND-DISCUSSIONS}, we discuss how the Fermi energy affects the plasmon-phonon coupling and, finally, in Sec.~\ref{sec:conclusion}, we present our conclusions.

\section{Plasmon-phonon-polaritons and hybrid modes \label{sec:THEORETICAL-FRAMEWORK}}
Dirac plasmons, density oscillations of Dirac fermions in graphene, can be obtained from the total system's dielectric function $\epsilon(q,\omega)$ within the random phase approximation (RPA)~\citep{GabrieleGiuliani2008,Fetter2003}. To do so, we find the solution of the plasmon equation which corresponds to the zeroes of $\epsilon(q,\omega)$ taking~\citep{GabrieleGiuliani2008,Fetter2003,Maier2007,Hwang2007,Wunsch2006,Principi2009}
\begin{equation}
    \epsilon(q,\omega) = 1 - v(q,\omega) \tilde{\chi}_{\rm nn}(q,\omega) = 0~,
    \label{plasmon_equation}
\end{equation}
where $v\left(q,\omega\right)$ is the Fourier transform of the Coulomb interaction between the Dirac electrons and $\tilde{\chi}_{\rm nn}(q,\omega)$ is the proper density-density response function~\cite{GabrieleGiuliani2008}. In general, both functions depend on the properties of the system as a whole. Nevertheless, within the RPA, we can approximate $\tilde{\chi}_{\rm nn}(q,\omega)$ by the non-interacting density-density response function of a 2D massless Dirac fermion $\chi^{0}(q,\omega)$, which depends only on the properties of graphene~\cite{Hwang2007,Wunsch2006, Principi2009}. On the other hand, $v\left(q,\omega\right)$ describes the electromagnetic field lines that mainly propagate through the surrounding of the graphene sheet, and are, therefore, strongly affected by them. In general, the 2D Fourier transform of the electron-electron Coulomb interaction is defined as
\begin{equation}
   v(q,\omega)=\frac{2\pi e^{2}}{q\epsilon_{\text{env}}(q,d)}~.
    \label{coulomb_interaction}
\end{equation}
As one can see from Eq.~(\ref{coulomb_interaction}), it is the screening of the Coulomb interaction introduced by the effective dynamical background dielectric function $\epsilon_{\text{env}}(q)$ that encodes the presence of the environment. To include the contribution of the TMDs thickness $d$ (see Fig.~\ref{Fig_vdWhs_plamon_phonon_coupling}(a)) to the screening, we define the background dielectric function as~\cite{Tomadin2015}
%
%\begin{equation}
%\epsilon_{\text{env}}(q)=\left(\frac{2}{\epsilon_{a}+\epsilon_{b}}\frac{\sqrt{\tilde{\epsilon}_{x}\tilde{\epsilon}_{z}}+\epsilon_{b}\hspace{0.1cm}\text{tanh\ensuremath{\left(qd\sqrt{\tild%e{\epsilon}_{x}/\tilde{\epsilon}_{z}}\right)}}}{\sqrt{\tilde{\epsilon}_{x}\tilde{\epsilon}_{z}}+\tilde{\epsilon}\hspace{0.1cm}\text{tanh\ensuremath{\left(qd\sqrt{\tilde{\epsilon}_{x}/\til%de{\epsilon}_{z}}\right)}}}\right)^{-1}~.
%\label{eq:new_eps}
%\end{equation}
%
\begin{equation}
\epsilon_{\text{env}}(q,d)=\left(\frac{2}{\epsilon_{a}+\epsilon_{b}}
\frac{\sqrt{\epsilon_{x}(d)\epsilon_{z}(d)}+\epsilon_{b}\hspace{0.1cm}\xi(d)}{\sqrt{\epsilon_{x}(d)\epsilon_{z}(d)}+\tilde{\epsilon}\hspace{0.1cm}\xi(d)}\right)^{-1}~.
\label{eq:new_eps}
\end{equation}
In Eq.~(\ref{eq:new_eps}), we have $\xi(d)=\text{tanh}(qd\sqrt{\epsilon_{x}(d)/\epsilon_{z}(d)}$ and $\tilde{\varepsilon}=\left(\epsilon_{x}(d)\epsilon_{z}(d)+\epsilon_{a}\epsilon_{b})\right/(\epsilon_{a}+\epsilon_{b})$. $\epsilon_{a,b}=1$ is the dielectric constant of the vacuum above and below the 2D materials slab. $\epsilon_x(d)$ and $\epsilon_z(d)$ are, respectively, the static in-plane and out-of-plane dielectric constants of the TMDs, where we have modiﬁed the notation to explicitly indicate its dependence on the TMDs thickness~\cite{Laturia2018,Cavalcante2018}. In order to facilitate the understanding of how plasmons couple with phonons, giving rise to hybrid modes, we assume that the plasmon dispersion attains its long-wavelength form,~\citep{Hwang2007,Goncalves2015,Wunsch2006}
\begin{equation}
\hbar\omega_{pl}=\sqrt{\frac{\alpha_{ee}N_{F}\hbar v_{F}}{2}\frac{E_{F}q}{\epsilon_{\text{env}}(q,d)}}~.
\label{eq. longwave}
\end{equation}
In Eq.~(\ref{eq. longwave}), $\alpha_{\rm ee}=2.2$, $N_{F} = 4$ and $v_{\rm F} = 10^{6}$~m/s are parameters related to the graphene sheet corresponding to the graphene fine structure constant, the number of Fermion flavours and the Fermi velocity, respectively~\cite{Neto2009}. $E_{\rm F}$ is the Fermi level of graphene.

\subsection{Coupling Dirac plasmon to phonons polaritons}\label{SubSec: Coupling plasmons with substrate phonons}
To introduce the concept of plasmon-phonon coupling, the simple classical analogy with two coupled harmonic oscillators, pictorially represented in Fig.~\ref{Fig_vdWhs_plamon_phonon_coupling}(c) with ``plasmon" and ``phonon" representing the masses $a$ and $b$, respectively, is commonly used~\cite{Hertzog2019,torma2014strong}. When $\kappa \neq 0$ the two oscillators interact with each other, forming a unique system, with hybridized eigenfrequencies~\cite{Novotny2010}. Due to this hybridization, an anticrossing of dispersion curves is formed, resulting in a coupling strength\cite{Novotny2010}:
\begin{equation}
2\Omega = \frac{\kappa}{\sqrt{{m_a \omega_a   m_b \omega_b}}}~.
\label{Eq: novotony}
\end{equation}

In the context of SPPPs, the coupling is similar to this classical point of view: when Dirac plasmons couple to the TMDs IR-active phonons, a hybridization occurs at $\omega_{\rm pl}=\omega_{\rm ph}$, giving rise to an anticrossing in the SPPPs dispersion for frequencies close to the phonon frequency, as presented in Fig.~\ref{Fig_vdWhs_plamon_phonon_coupling}(d). For frequencies further away from the phonon frequency, the original energy remains practically unchanged from the uncoupled case. In other words, the uncoupled phonon ($\omega_{\text{ph}}$) and graphene plasmon ($\omega_{\rm pl}$) frequencies, represented in Fig.~\ref{Fig_vdWhs_plamon_phonon_coupling}(d) as a horizontal green dashed and a solid red ($\propto \sqrt{q}$) lines, respectively, presents hybrid modes ($\omega_{+}$ and $\omega_{-}$) close the phonon frequency when coupled. 

To quantify the SPPPs coupling ($\Omega$), we start from its Hamiltonian, defined as~\cite{Ribeiro2020}
\begin{equation}
    H=H_{\rm pl}+H_{\rm ph}+H_{\rm pl-ph}~.
\end{equation}
Here, $H_{\rm pl}$ is the Hamiltonian for the plasmons in the absence of the coupling to the phonons $H_{\rm ph}$, while $H_{\rm pl-ph}$ describes the coupling between them. In second quantization notation, this yields~\cite{Ribeiro2020}
\begin{equation}
    H=\hbar[\omega_{\rm pl} \hat{a}^{\dagger}_{\bf{q}} \hat{a}_{\bf{q}}+\omega_{\rm ph}\hat{b}^{\dagger}_{{\bf{q}}}\hat{b}_{{\bf{q}}}+\Omega_{\bf{q}}(\hat{a}^{\dagger}_{\bf{q}}+\hat{a}_{{\bf{-q}}})(\hat{b}^{\dagger}_{{\bf{-q}}}+\hat{b}_{\bf{q}})]~,
\label{eq. second quantization}
\end{equation}
where $\hat{a}^{\dagger}_{\bf{q}}$ and $\hat{a}_{\bf{q}}$ are creation and annihilation operators, respectively, for a Dirac plasmon (SP$^2$) with frequency $\hbar\omega_{\rm pl}$ given by Eq.~(\ref{eq. longwave}), and wave vector $\textbf{q}$. $\hat{b}^{\dagger}_{\bf{q}}$ and $\hat{b}_{\bf{q}}$ are those for the collective vibration modes with energy $\hbar\omega_{\rm ph}$ (taken as a constant, as presented in Tab.~\ref{Tab: MX2-AppendixA}). In Eq.~(\ref{eq. second quantization}), $\Omega_{\bf{q}}$ plays the role of the coupling energy associated with the interaction between phonons and the Dirac plasmon. Consequently, the eigenfrequencies are obtained taking $\det[H]=0$, resulting in~\cite{Ribeiro2020}
\begin{equation}
\omega_{\pm}^2\hspace{-1 mm}=\hspace{-1 mm}\frac{1}{2}\hspace{-1 mm}\left[\omega_{\rm ph}^2
+\omega_{\rm pl}^2\pm\sqrt{(\omega_{\rm ph}^2-\omega_{\rm pl}^2)^{2}+16\Omega^{2}\omega_{\rm ph}\omega_{\rm pl}}\right].
\label{eq:eigenfrequencies}
\end{equation}
Equation (\ref{eq:eigenfrequencies}) is similar to those obtained from a classical system formed by two coupled oscillators~\citep{Novotny2010}, where the coupling $\Omega$ arises due to the hybridization between two (quasi)-particles, as shown in Fig.~\ref{Fig_vdWhs_plamon_phonon_coupling}(d).

The goal of the current study is to identify the coupling strength $\Omega$ from realistic calculations of the anticrossing between plasmon and phonon branches. From Eq.~(\ref{eq:eigenfrequencies}), one finds that $\Omega$ can be calculated in two ways: on the one hand, one can find the minimum of the energy difference between the two branches, i.e. $\Omega_{\rm min}=\min_{q}(\omega_{+}(q)-\omega_{-}(q))$ (see Fig.~\ref{Fig_vdWhs_plamon_phonon_coupling}(d)). On the other hand, it can also be calculated at the crossing point of the phonon frequency with the unperturbed plasmon. Here, the coupling strength corresponds to the energy difference between the two branches evaluated at the wave vector $q_{\rm pl}(\omega_{\rm ph})$, i.e. $\Omega_{\rm cp} = \omega_{+}(q_{\rm pl})-\omega_{-}(q_{\rm pl})$. Note that in the case of a system consisting of a single plasmon and phonon, both methods are equivalent, because in that case Eq.~(\ref{eq. second quantization}) corresponds to the full system. However, once multiple phonons start to interfere with the plasmon, the model is only approximately correct and both methods will not yield the same result. In order to quantify the plasmon-phonon interaction also in the presence of multiple phonons, we always evaluate $\Omega$ using both methods. If the difference between both methods is large with respect to the nominal value of the coupling, i.e. if $\Delta \Omega = \vert \Omega_{\rm cp} - \Omega_{min} \vert \sim \Omega_{i}$, a hierarchy is necessary. For example, in the case where there are two relevant phonon modes, as discussed in the succeeding examples of this work, we find that it is necessary to calculate  $\Omega_{\rm min}$ for the smallest value, while $\Omega_{\rm cp}$ is needed for the strongest coupling. This is because, in that case, the plasmon-phonon coupling becomes of the order of the frequency difference between the two involved phonon modes.

\begin{table}[!t]
\centering{}
\caption{Phonon frequencies for the free-standing monolayer of $\text{MoS}_{2}$ and $\text{WS}_{2}$ considered in the QEH calculations. Their vibrational phonon modes are represented by $\text{E}^{\prime\prime}$ (R), $\text{E}^{\prime}$ (IR and R), $\text{A}^{\prime}_1$ (R) and $\text{A}^{\prime\prime}_2$ (IR), where IR (R) means that the mode is active for infrared (Raman) excitations~ \cite{Zhang2015a,Zhao2013,MolinaSanchez2011,Peng2016,Berkdemir2013,Sengupta2015}.}

\begin{tabular}{p{1cm} p{0.25cm} p{1.5cm} p{1.5cm} p{1.5cm} p{1.5cm}}
\hline\hline

&&   \multicolumn{4}{c}{Phonon frequencies (meV)}  \\ 
\hline
&& 1 ($E''$) & 2 ($E'$) & 3 ($A'_{1}$) & 4 ($A''_{2}$)\\
\hline
MoS$_{2}$ && 34.19 & 46.35 & 47.59 & 56.80 \\
\hline
WS$_{2}$ && 35.56 & 42.85 & 50.12 & 52.98 \\
%\hline
%MoSe$_{2}$ && 20.18 & 28.10 & 34.37 & 42.53 \\
%\hline
%WSe$_{2}$ && 20.71 & 29.67 & 30.19 & 37.21 \\
\hline\hline
\end{tabular}
\label{Tab: MX2-AppendixA}
\end{table}

\subsection{Quantum electrostatic heterostructure}
To obtain realistic results for the plasmon-phonon coupling, we used a DFT-based method known as the quantum-electrostatic heterostructure (QEH) model~\cite{Andersen2015}. This model has been demonstrated to be a very useful tool for the study of plasmons in different heterostructures~\citep{Lavor2020,cavalcante2019,Shirodkar2018,Nerl2017,Gjerding2020}. In the QEH model, the dielectric constant of the monolayer that composes the vdWhs is calculated individually within the DFT. Then, using Coulomb interaction, the contributions of each freestanding atomic layer are coupled, and the total responses of the vdWhs is obtained~\cite{Andersen2015}. The SPPPs coupling $\Omega$ is obtained from the loss function, which is defined as
\begin{equation}
    L(q,\omega) = -{\rm Im}\left[\frac{1}{\epsilon(q,\omega)}\right]~.
    \label{equation_lossfunction}
\end{equation}
The major advantage of the use of the QEH model is its database containing the dielectric building blocks of a large collection of 2D materials~\citep{Link1}, allowing us to reuse previously obtained DFT results. This enables the careful study of different vdWh systems on a layer-by-layer basis, without the need to treat the dielectric environment as slabs of bulk material. 

\section{Strength of plasmon-phonon coupling in van der Waals heterostructures\label{sec:RESULTS-AND-DISCUSSIONS}}
TMDs are slightly polar materials i.e its crystalline structure contains atoms with diﬀerent electronegativities, consequently, certain IR-active phonon modes at the $\Gamma$-point give rise to a macroscopic electric ﬁeld ~\cite{MolinaSanchez2011,Griffiths2017}. Both MoS$_2$ and WS$_2$, the TMDs considered in this paper, present four phonon modes labelled, in ascending order of energy (see Tab.~\ref{Tab: MX2-AppendixA}), as: $E^{\prime\prime}$ (R), $E^{\prime}$ (IR and R), $A^{\prime}_1$ (R) and $A^{\prime\prime}_2$ (IR), where IR (R) means that the phonon mode is active for infrared (Raman) excitations~\cite{Zhang2015a}. 

\begin{figure}[!t]
\centering{}\includegraphics[width=1\columnwidth]{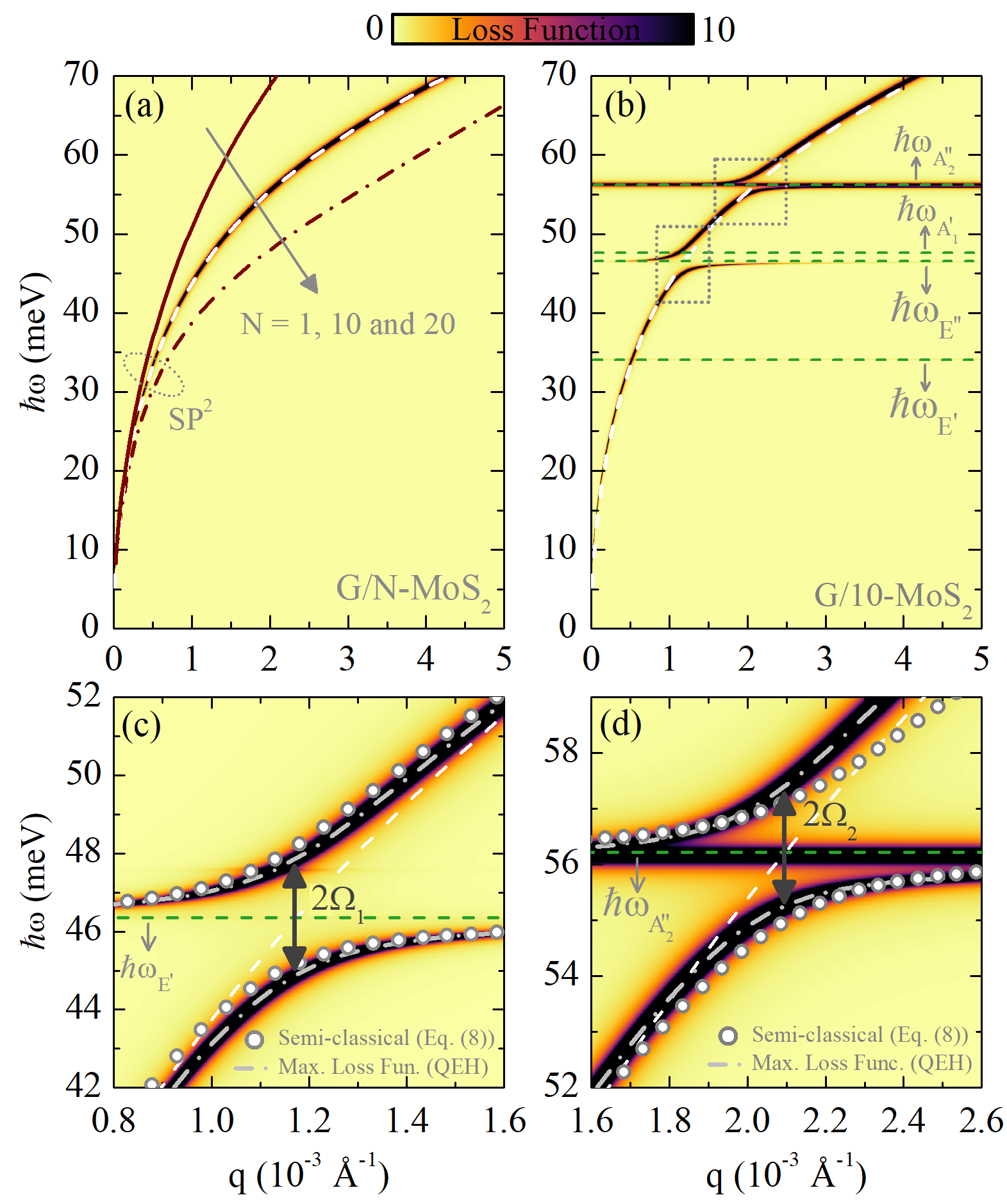}
\caption{(Color online) (a) Plasmon dispersion of the SP$^2$ for G/N-MoS$_2$ with $N$=1 (solid), 10 (white dashed) and 20 (dashed-dotted) at $E_{F}=100~$meV obtained from the QEH without plasmon-phonon coupling ($\Omega=0$). The loss function is shown as a color map for $N$ = 10. (b) SPPPs dispersion for G/10-MoS$_2$ with $E_F=100~$meV. The two regions with IR-active phonons modes, namely E$^{\prime\prime}$ and A$^{\prime\prime}_2$, that hybridize with the Dirac plasmons giving rise to anti-crossings in the eigenfrequencies when $\omega_{pl}=\omega_{ph}$, are highlighted by two rectangles. Horizontal green dashed lines represents the phonon frequencies (see Tab.~\ref{Tab: MX2-AppendixA}). (c) and (d) are magnifications of the results in (b) around the anticrossings, close to the $E^{\prime\prime}$ and A$^{\prime\prime}_2$ phonon modes, with frequencies $\hbar\omega_{E^{\prime}}$ and $\hbar\omega_{A^{\prime\prime}_{2}}$, respectively. In panel (c) and (d) $\Omega_{1(2)}$ represents the coupling strength between Dirac plasmon and IR-active in-plane (out-of-plane) vibrational phonon mode. Symbols are the eigenfrequencies obtained from the semi-classical model, Eq.~(\ref{eq:eigenfrequencies}). Dashed-doted gray lines are the maxima in the loss function, while the dashed white line is the SP$^2$ dispersion for reference.}
\label{Fig: SP2_dispersion}
\end{figure}

In Fig.~\ref{Fig: SP2_dispersion}(a), we present the plasmon dispersion of SP$^2$ modes, i.e Dirac plasmons with the surrounding polarization cloud~\cite{Griffiths2017, Goncalves2015}, but disregarding the TMDs phonon vibrations, at the Fermi energy given by $E_F = 100$~meV, for a  G/N-MoS$_2$ vdWhs, with $N=1$, 10 and 20 TMD layers. The loss functions obtained by the QEH calculation, shown as a color map for $N$ = 10, are in accordance with Eq.~(\ref{eq. longwave}), whose results are represented by white dashed curves in Fig.~\ref{Fig: SP2_dispersion}(a). As the number of layers increases, $q$ increases for a fixed frequency in the plasmon dispersion, since the total dielectric function of the environment $\epsilon_{\rm env}(q,d)$ also increases, since the screening is proportional to the number of layers. This is verified by the solid and dashed-dotted lines in Fig.~\ref{Fig: SP2_dispersion}(a), which represent the maxima of the loss function for $N = 1$ and 20, respectively. When phonon contributions are taken into account, as shown in Fig.~\ref{Fig: SP2_dispersion}(b), anticrossings in the SP$^2$ dispersion arise close to the regions where $\omega_{\rm pl}=\omega_{\rm ph}$. Although MoS$_2$ has four phonon modes, only two of them are IR-active, as mentioned earlier, giving rise to significant hybrid modes. These hybrid SPPPs modes are presented in Figs.~\ref{Fig: SP2_dispersion}(c)-(d) as a magnification of the two square boxes highlighted in Fig.~\ref{Fig: SP2_dispersion}(b). The coupling strength between the Dirac plasmons and the in-plane E$^\prime$ (out-of-plane A$^{\prime\prime}_{2}$) phonon mode is defined as $\Omega_1$ ($\Omega_2$). In panels (c) and (d), the symbols refer to the hybrid eigenfrequencies obtained from Eq.~(\ref{eq:eigenfrequencies}).

\subsection{The influence of the number of TMDs layers}
\begin{figure}[!b]
\centering{}\includegraphics[width=1\columnwidth]{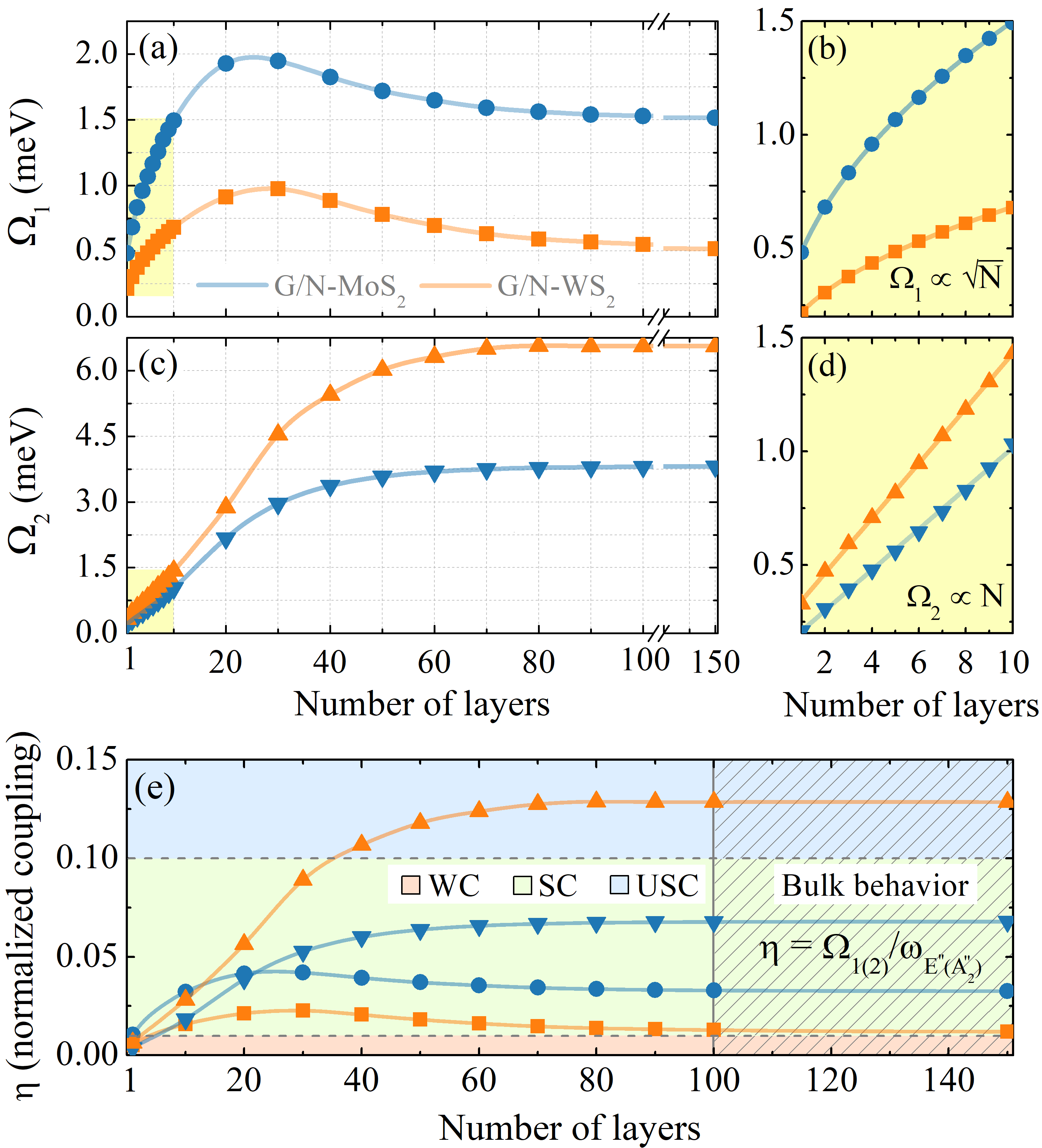}
\caption{(Color online) SPPPs coupling strength ($\Omega_{1(2)}$) as a function of the number of  TMDs layers for graphene at $E_F=100~$meV on top of N-MoS$_2$ (blue lines and circles) and N-WS$_2$ (orange lines and circles). (a) Coupling energy between Dirac plasmon and the IR-active in-plane $E^{\prime}$ phonon mode and (b) a magnification of the results in (a) from 1 to 10 layers (yellow region), emphasizing that $\Omega_1\propto\sqrt{N}$. (c) The same as in (a), but now for the coupling strength $\Omega_2$, i.e considering the IR-active out-of-plane A$^{\prime\prime}_2$ phonon mode. (d) Magnification in panel (c) from 1 to 10 layers (yellow region), emphasizing that $\Omega_2 \propto N$. (e) SPPPs coupling strength normalized in relation to their respective phonon frequencies defined as $\eta = \Omega_{1(2)} / \omega_{E^{\prime\prime}(A^{\prime\prime}_2)}$. Three different regions, blue,  green  and  pink, represent the WC ($\eta < 0.01$), SC ($0.01   \leq \eta < 0.1$) and USC ($\eta\geq0.1$), respectively~\cite{kockum2019ultrastrong}. The hatched area represents the bulk limit of the SPPPs coupling, reached for approximately 100 TMDs layers.}
\label{Fig: Plasmon_phonon_strenght}
\end{figure}
Using the QEH, we show in Fig.~\ref{Fig: Plasmon_phonon_strenght}(a)-(d) the evolution of the SPPPs coupling strength ($\Omega_{1(2)}$) as a function of the number of layers for a vdWhs composed by MLG on top of $N$-MoS$_2$ (blue symbols and lines) and $N$-WS$_2$ (orange symbols and lines). As the number of layers increase, the SPPPs coupling ($\Omega_{1(2)}$) also increases, since more oscillators are involved, i.e more phonons are available to couple with the Dirac plasmons~\cite{jia2015tunable,huck2016strong}. For a few TMD layers ($N<10$), there are two important and peculiar aspects to be considered in here: (i) the $\sqrt{N}$ behavior of $\Omega_1$, that is the coupling between plasmon and in-plane phonon modes, Fig.~\ref{Fig: Plasmon_phonon_strenght}(b), and (ii) the linear dependence of the out-of-plane phonon coupling $\Omega_2$, Fig.~\ref{Fig: Plasmon_phonon_strenght}(d). To explain this behaviour, we analyse the effective dielectric function~\cite{low2014novel,jia2015tunable}
\begin{equation}
    \epsilon^\text{eff}_{i}  \approx \epsilon_{\text{env}} \qty(1 - \frac{\omega^{2}_{pl} } {\omega^2} - \frac{M_{i}\delta^2_{i}}{\omega^2 -\omega^{2}_{\rm ph,\textit{i}}+\delta^2_{i}})~.
    \label{Eq: dielectric constant Tony}
\end{equation}
Equation ~(\ref{Eq: dielectric constant Tony}) describes an effective coupling between a plasmon and the $i$-th phonon mode. Notice that in the current case, the $E^{\prime}$ phonon and the $A_2^{\prime\prime}$ phonons exhibit different geometric properties. The former is an in-plane mode of which degeneracy $M_i$ increases linearly with the number of layers $N$. Conversely, the latter is an out-of-plane mode with degeneracy scaling with $N^{2}$. In Eq.~(\ref{Eq: dielectric constant Tony}), $\delta_{i}$ is the coupling between a single TMD layer and the Dirac plasmon. Notice that this approximation only holds as long as the penetration depth of the plasmon mode is larger than the TMD thickness. In this case, the zeroes of Eq.~(\ref{Eq: dielectric constant Tony}) yield the relation between the hybrid modes as~\cite{jia2015tunable}
\begin{equation}
\omega_{i}  \approx  \omega_{\rm ph,\textit{i}} \pm \frac{1}{2}\sqrt{M_{i}}\delta_{i}~.
\label{Eq: layer dependence}
\end{equation}
Therefore, Eq.~(\ref{Eq: layer dependence}) reveals that, within this model, the SPPPs coupling $\Omega_{1(2)}$ is indeed expected to depend on the number of layers $N$ as $\Omega_{1(2)}=\sqrt{M_{1(2)}}\delta_{1(2)}$, where $M_{1(2)} = N^{1(2)}$. 

\subsection{SPPPs interaction: weak, strong and ultra-strong coupling regime}
We now define the normalized parameter $\eta = \Omega_{1(2)} / \omega_{E^{\prime\prime}(A^{\prime\prime}_2)}$~\cite{kockum2019ultrastrong} as a way to quantify the coupling strength. Figure \ref{Fig: Plasmon_phonon_strenght}(e) shows the normalized SPPPs coupling $\eta$ as a function of the number of $N$-MoS$_2$ and $N$-WS$_2$ layers. Three different regions, blue,  green  and  pink, represent the WC ($\eta < 0.01$), SC ($0.01 \leq \eta < 0.1$) and USC ($\eta\geq 0.1$) regimes, respectively~\cite{kockum2019ultrastrong}. A remarkable result is obtained for the coupling between Dirac plasmons and the IR-active out-of-plane WS$_2$ phonon mode, where we observe that they reach the USC regime, as illustrated in Fig.~\ref{Fig: Plasmon_phonon_strenght}(e) by orange triangles. Furthermore, for $N>100$ all results remain unchanged, showing that the bulk behavior was reached for 100 TMD layers or more (see hatched area in Fig.~\ref{Fig: Plasmon_phonon_strenght}(e)).

\begin{figure}[!t]
\centering{}\includegraphics[width=0.95\columnwidth]{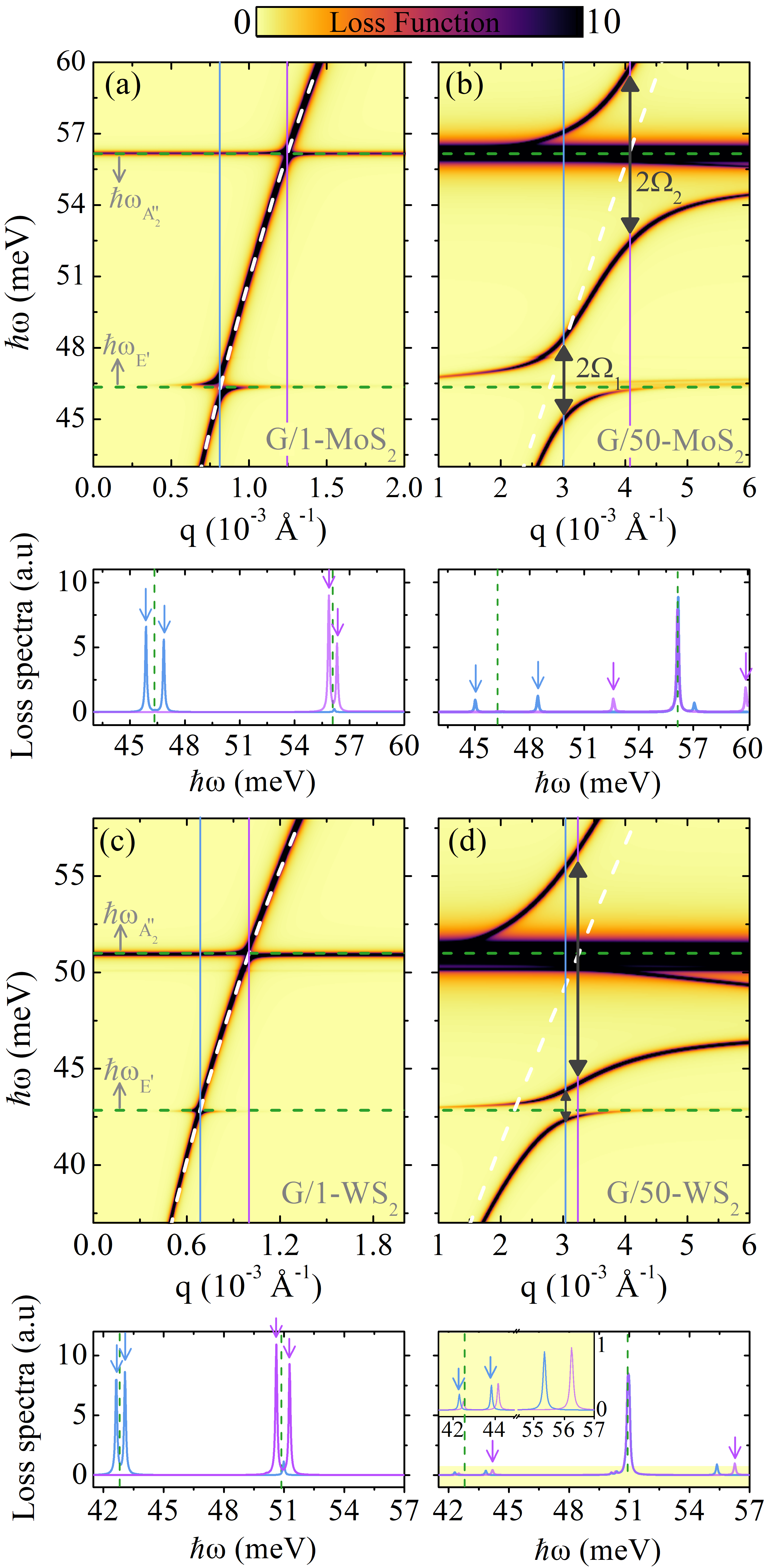}
\caption{(Color online) Overview of SPPPs dispersion in the (q,$\omega$)-plane through the loss function for a MLG, at $E_F=100~$meV, on top of (a) 1 and (b) 50 MoS$_2$, and on top of (c) 1 and (d) 50 WS$_2$. $\Omega_1$ ($\Omega_2$) corresponds to the coupling strength between Dirac plasmons and the IR-active in-plane E$^{\prime}$ (out-of-plane A$^{\prime\prime}_2$) phonon mode. The horizontal green curves correspond to the uncoupled phonon modes calculated for a monolayer of each correspondent TMD (see Tab.~\ref{Tab: MX2-AppendixA} for the corespondent phonon frequencies $\hbar\omega_{E^{\prime}}$ and $\hbar\omega_{A^{\prime\prime}_2}$). The uncoupled SP$^2$ plasmons are represented by white dashed lines, for reference. The results in each bottom panel are the loss spectra for a fixed $q$ at the point were the SPPPs coupling strengths $\Omega_{1(2)}$ were calculated. In the bottom panel (d), a magnification of the loss spectra is shown as inset.}
\label{Fig: Dispersion_MLG_MX2_ALL}
\end{figure}

To illustrate the WC, SC and USC regime in a TMDs-based vdWhs, we shown in Fig.~\ref{Fig: Dispersion_MLG_MX2_ALL} an overview of SPPPs dispersion in the (q,$\omega$)-plane through the color maps of the loss function, defined by Eq.~(\ref{equation_lossfunction}), and the loss spectra for a fixed $q$ at the point were the SPPPs coupling $\Omega_{1(2)}$ were calculated. As expected, for a MLG on top of 1-MoS$_2$ or 1-WS$_2$, Figs.~\ref{Fig: Dispersion_MLG_MX2_ALL}(a) and ~\ref{Fig: Dispersion_MLG_MX2_ALL}(c), respectively, the SPPPs coupling are in the WC regime. In this case, the modes that compose the anticrossing, arising due their hybridization, are practically indistinguishable, as compared to the line width of the non-coupled modes. The loss spectra below each panel emphasizes how weak this couplings is, since the peaks, represented by blue (purple) arrows for $\Omega_1$ ($\Omega_2$), are very close to each other, presenting a normalized coupling $\eta$ less than 0.01. In Fig.~\ref{Fig: Dispersion_MLG_MX2_ALL}(b), both $\Omega_1$ and $\Omega_2$ are in the SC, presenting a well defined anticrossing and a loss spectra with well separated peaks, where $\eta$ is given by 0.047 and 0.063, respectively. Finally, although $\Omega_1$ in Fig.~\ref{Fig: Dispersion_MLG_MX2_ALL}(d) presents a SC, with $\eta=0.18$, $\Omega_2$ is in the USC coupling regime with $\eta=0.12$ in this case.
\vspace{-5mm}
\subsection{Tuning the SPPPs coupling strength through the Fermi energy}\label{subsec:Omega_vs_Fermi_Energy}
\begin{figure}[!t]
\centering{}\includegraphics[width=1\columnwidth]{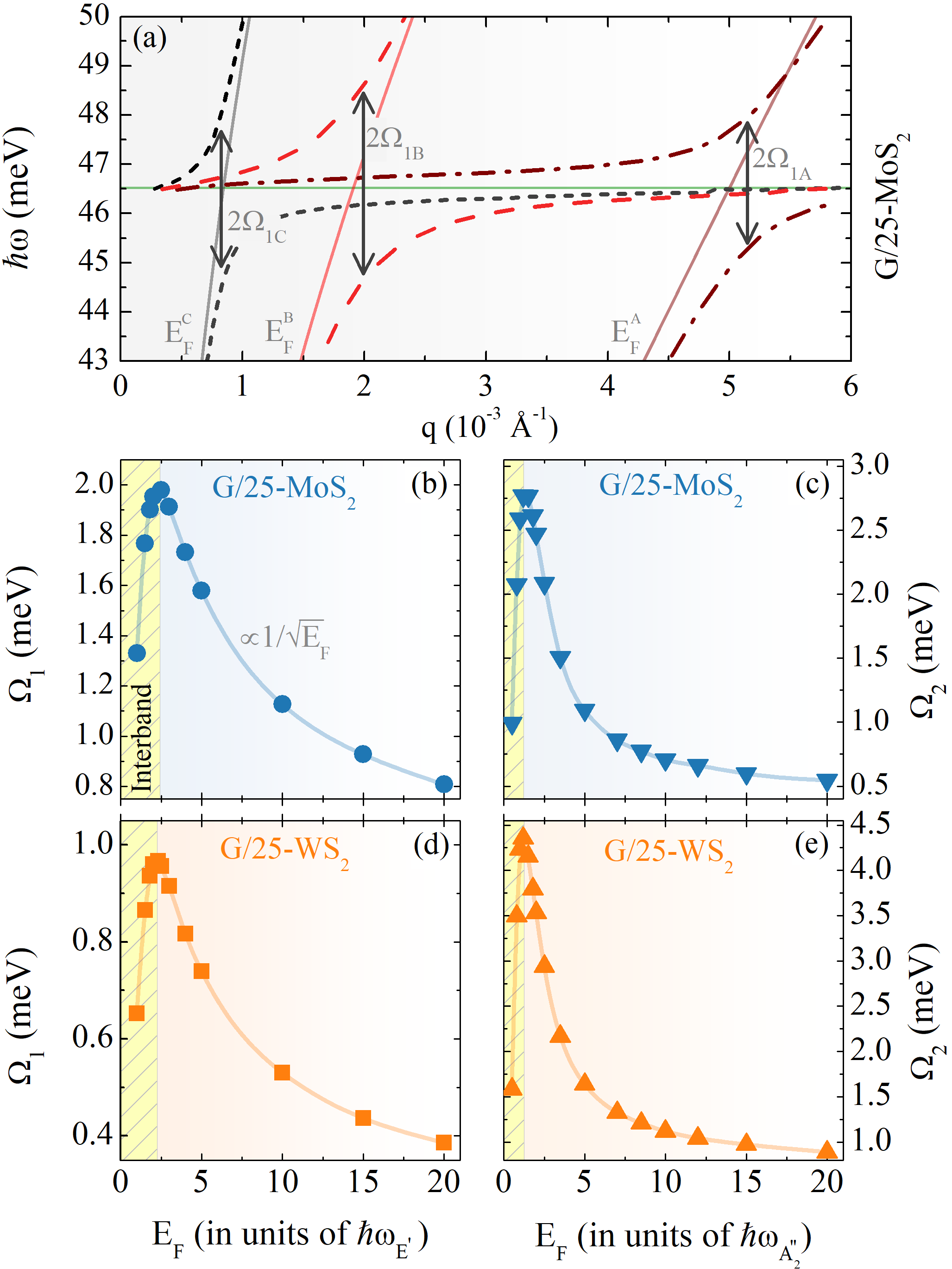}
\caption{(Color online) Tuning the plasmon-phonon coupling strength $\Omega_{1(2)}$ by changing the Fermi energy (in units of the corresponding phonon frequency). (a) Plasmonic dispersion of G/25-MoS$_2$ for different values of the Fermi energy (in units of  $\hbar\omega_{E^{\prime}}$) given by $E_F^A=1\hbar\omega_{E^{\prime}}$, $E_F^B=2.3\hbar\omega_{E^{\prime}}$ and $E_F^C=3.8\hbar\omega_{E^{\prime}}$. The uncoupled phonon state corresponds to the horizontal solid green line and the SP$_2$ plasmons are represented by the square root ($\propto \sqrt{q}$) solid lines, for reference. The SPPPs couplings (b)-(c) $\Omega_1$ and (d)-(e) $\Omega_2$ are shown as a function of the Fermi energy for G/25-MoS$_2$. The yellow region in (b)-(e) represents the interband regime, where the plasmon dispersion is damped. After that, $\Omega_{1(2)}  \propto 1/\sqrt{E_F}$, i.e the Fermi energy is large enough to keep the plasmon-phonon dispersion in the long-wavelength limit, keeping the plasmonic dispersion below the interband region.}
\label{Fig: Omega1_omega2_vs_fermi_energy}
\end{figure}
Figure~\ref{Fig: Omega1_omega2_vs_fermi_energy} shows how the Fermi energy can be use to tune the SPPPs coupling, as to maximize the plasmon-phonon interaction. In Fig.~\ref{Fig: Omega1_omega2_vs_fermi_energy}(a), we present the SPPPs dispersion for a vdWhs made by G/25-MoS$_2$ for three different values of the Fermi energy (in units of the phonon frequency $\hbar\omega_{E^{\prime}}$, see Tab.~\ref{Tab: MX2-AppendixA}): $E_F^A=1\hbar\omega_{E^{\prime}}$, $E_F^B=2.3\hbar\omega_{E^{\prime}}$ and $E_F^C=3.8\hbar\omega_{E^{\prime}}$, represented by the black dotted, red dashed and brown dash-dotted lines, respectively. The horizontal green line is the phonon frequency and the other solid lines are the SP$_2$ dispersion for reference. Fig.~\ref{Fig: Omega1_omega2_vs_fermi_energy}(a) shows that there is a Fermi energy value that maximizes the SPPPs coupling strength. To explain this, we show in Figs.~\ref{Fig: Omega1_omega2_vs_fermi_energy}(b)-(e) the SPPPs coupling parameters $\Omega_1$ and $\Omega_2$ as a function of the Fermi energy. In all situations, $\Omega_1$ and $\Omega_2$ increase until they reach a maximum value, and then they decrease with $E_F$, exhibiting $\propto 1/\sqrt{E_F}$ dependence. 

To explain this behaviour, we identify two different coupling mechanisms that depend on the Fermi energy $E_{\rm F}$. If the $E_{\rm F}$ is large, due to Pauli blocking, single-particle inter-band processes are suppressed. In that case, the Dirac liquid effectively behaves as a liquid of Fermions with a mass equal to the cyclotron mass $m_{\rm c} = 2 E_{\rm F} /v^{2}_{\rm F}$~\cite{Neto2009}. Eq.~(\ref{Eq: novotony}) shows that in this case the plasmon-phonon coupling $\Omega$ is expected to decrease as $1/\sqrt{E_F}$. However, when the Fermi energy is small, Pauli blocking is lifted and inter-band single-particle processes are allowed~\cite{Low2014,Goncalves2015}. This strongly inhibits plasmon lifetime and, therefore, suppresses plasmon-phonon coupling.

Note that, for the vdWhs considered in Figs.~\ref{Fig: Omega1_omega2_vs_fermi_energy}(b)-(e), both SPPPs coupling $\Omega_1$ and $\Omega_2$ are in the SC regime. However, controlling the Fermi energy and increasing the number of layers it is possible to go from the SC to even the USC regime. The latter can be reached for $\Omega_2$ in a MLG on top of 50 (or more) WS$_2$ layers, for example.

\section{Conclusions}\label{sec:conclusion}

We have demonstrated how graphene (Dirac) plasmons couple to IR-active in-plane $E^{\prime}$ and out-of-plane $A^{\prime\prime}_{2}$ phonon modes in transition metal dichalcogenide-based van der Waals heterostructures, from few layers until the bulk limit. In order to do so, we have presented a semi-classical theory, obtained from the random phase approximation, to calculate the surface plasmon-phonon polaritons dispersion in the q-$\omega$ plane. Comparing this semi-classical theory to the results obtained through a DFT-based method, known as the quantum-electrostatic heterostructure, we have shown that the semi-classical approach provides an excellent match for many TMDs layers, capturing all relevant characteristics of the surface plasmon-phonon polaritons. 

Furthermore, using the quantum-electrostatic heterostructure model, we have calculated the loss function of vdWHs composed by monolayer graphene on top of TMDs multi-layers. Our results prove that, although we have weak and strong coupling regimes in this TMDs-based vdWhs, it is also possible to achieve the ultra strong coupling regime for the coupling between Dirac plasmons and $A^{\prime\prime}_2$ for 40 or more WS$_2$ layers. In addition, we explain the nature of the graphene plasmons coupling to IR-active $E^{\prime}$ and $A^{\prime\prime}_{2}$ phonon modes, from a few TMDs layers to the bulk behavior. Not less important, we have demonstrated the possibility of tune the SPPPs coupling strength through the graphene Fermi energy, explaining its $1/\sqrt{E_F}$ dependence. It is important to highlight that plasmons in graphene can be experimentally observed using, for example, scattering-type scanning near-field optical microscope (s-SNOM) in photocurrent mode. Therefore, using current experimental techniques, our results suggest the possibility of creating/exciting SPPPs and to study the coupling regimes discussed here for vdWhs composed by graphene and $\text{MoS}_{2}$ or $\text{WS}_{2}$.

\section*{ACKNOWLEDGMENTS}
Discussions with D. J. P. de Sousa and L. S. R. Cavalcante are gratefully acknowledged. This work was financially supported by the Brazilian Council for Research (CNPq), Brazilian National Council for the Improvement of Higher Education (CAPES) and by the Research Foundation Flanders (FWO), through postdoctoral fellowships granted to B.V.D and A.C. 
\appendix
\counterwithin{figure}{section}
\section{Plasmon-phonon coupling strength at the minimal energy splitting and at the crossing point}\label{App.A}

\begin{figure}[!t]
\centering{}\includegraphics[width=1\columnwidth]{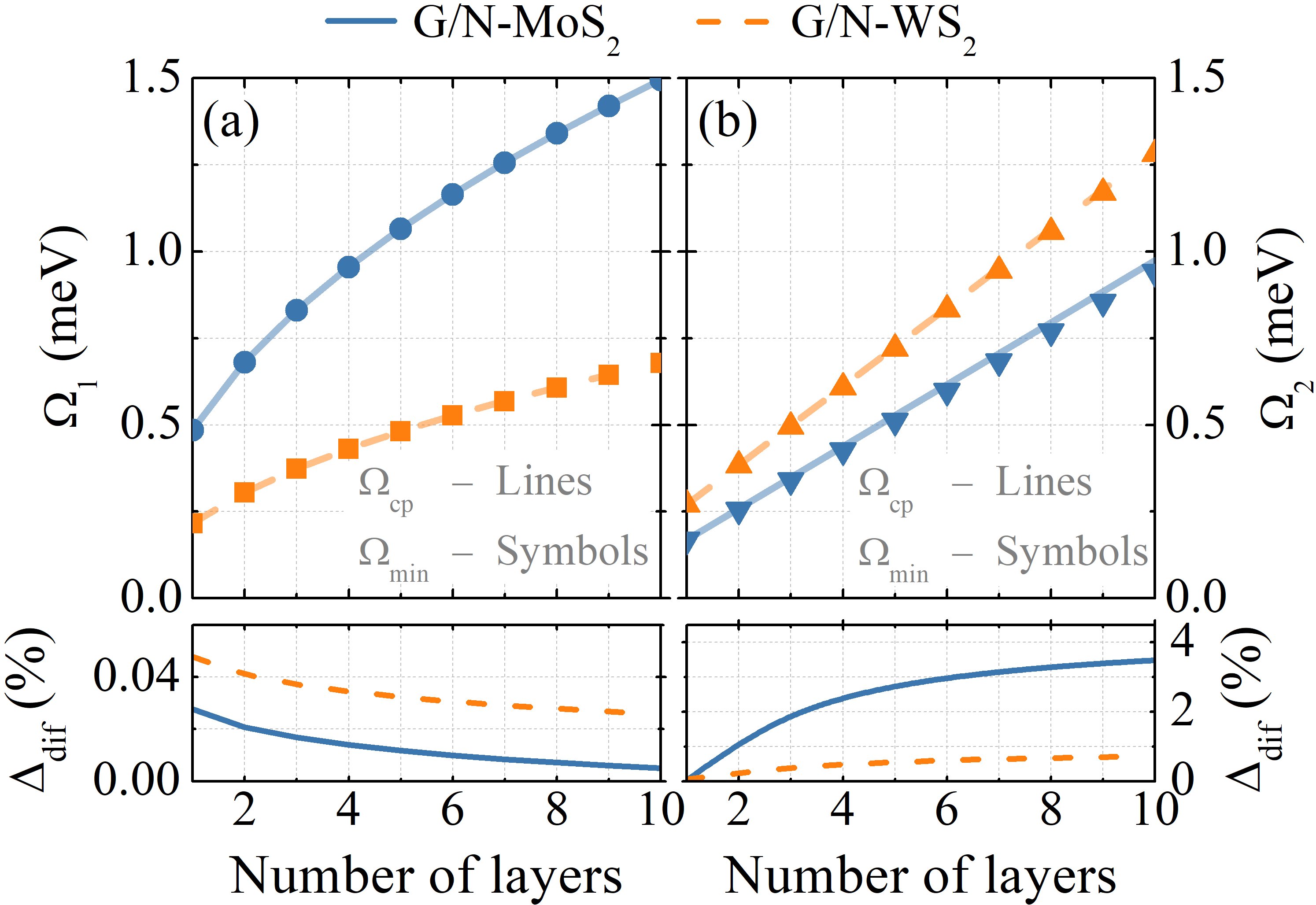}
\caption{(Color online) Comparison between the SPPPs coupling strength, for few layers of TMD, obtained from the minimal energy splitting thought $\Omega_{\rm min}$ (symbols) and at the crossing point $\Omega_{\rm cp}$ (lines). (a) Results of the comparison between plasmons and in-plane $E^{\prime}$ phonons in a vdWhs composed by G/N-MoS$_2$  (G/N-WS$_2$), blue (orange) symbols and lines, with $E_F=100~$meV. (b) The same as in (a), but now for the coupling between plasmons and out-of-plane $A^{\prime\prime}_2$ phonons. The bottom panels in (a) and (b) present the relative difference between both procedures, i.e $\Omega_{\rm min}$ and $\Omega_{\rm cp}$, defined as $\Delta_{\rm dif}=100\abs{\Omega_{\rm cp}-\Omega_{\rm min}}/\Omega_{\rm cp}$.}
\label{Fig: App.A}
\end{figure}
Here, we provide a comparison for the SPPPs coupling strength as obtained from the minimum of the energy difference  between  the  two  branches $\Omega_{\rm min}$ and those obtained at the crossing point $\Omega_{\rm cp}$, as previously discussed in Sec.~\ref{SubSec: Coupling plasmons with substrate phonons}. Results are depicted in Fig.~\ref{Fig: App.A}(a), for the coupling between Dirac plasmons and IR-active in-plane $E^{\prime}$ phonon mode ($\Omega_1$), and in Fig.~\ref{Fig: App.A}(b) for Dirac plasmons and IR-active out-of-plane $A^{\prime\prime}_{2}$ phonon mode ($\Omega_2$). Blue (orange) results in both panels are for G/N-MoS$_2$ (G/N-WS$_2$), with $N$ from 1 to 10 TMDs layers, while symbols (lines) represents the results obtained from $\Omega_{\rm min}$ ($\Omega_{\rm cp}$). Both methods yield practically the same results. To quantify the difference between them, we show in the bottom panels the relative difference $\Delta_{\rm dif}$ between the results from both methods, defined as $\Delta_{\rm dif}=100\abs{\Omega_{\rm cp}-\Omega_{\rm min}}/\Omega_{\rm cp}$. For $\Omega_1$, the bottom panel in Fig.~\ref{Fig: App.A}(a) shows relative differences lower than 0.1~\%, while for $\Omega_2$, in the bottom panel of Fig.~\ref{Fig: App.A}(b), they are less than 4~\%.

\bibliographystyle{apsrev4-2}
\bibliography{myreferences}

\end{document}